\documentclass[a4paper,11pt]{article}
\pdfoutput=1 % if your are submitting a pdflatex (i.e. if you have
\usepackage{jheppub} % for details on the use of the package, please
\usepackage[T1]{fontenc} % if needed
\usepackage{comment}
\usepackage{tikz}
\usepackage{amsmath,amssymb,amsfonts}
\usepackage{latexsym}
\usepackage{bbm}
\usepackage{amssymb}
\usepackage{dsfont}
\usepackage{slashed, tensor}
\usepackage{booktabs}
\usepackage{graphicx}
\usepackage{mathrsfs}

\newcommand{\be}{\begin{equation}}
	\newcommand{\ee}{\end{equation}}

\newcommand{\bea}{\begin{eqnarray}}
	\newcommand{\eea}{\end{eqnarray}}
\newcommand{\ba}{\begin{array}}
	\newcommand{\ea}{\end{array}}

\title{Manifest color-kinematics duality for point particles interacting with self-dual fields}

\author[a,b]{Vyacheslav Ivanovskiy}
\author[b,c]{and Dmitry Ponomarev}

\affiliation[a]{Moscow Institute of Physics and Technology,
  Dolgoprudny, 141701, Russia}
\affiliation[b]{Institute for Theoretical and Mathematical Physics,\\
Lomonosov Moscow State University,  Moscow, 119991, Russia}
\affiliation[c]{I.E. Tamm Theory Department, Lebedev Physical Institute,
 Moscow, 119991, Russia}

\emailAdd{ivanovskiy.va@phystech.edu}
\emailAdd{ponomarev@lpi.ru}

\abstract{We find that point particles interacting with a self-dual Yang-Mills field and self-dual gravity manifestly satisfy color-kinematics duality at the level of action.
In a similar way color-kinematics duality also holds for a scalar field minimally coupled to a self-dual Yang-Mills field and self-dual gravity. 
By applying the appropriate limiting procedure to these scalar field theories we reproduce point particle theories we started from. This allows us to connect worldline color-kinematics duality to  amplitude color-kinematics duality in field theory. Considering that point particles act as sources of classical solutions, our results may be regarded as a step towards establishing a precise relation between the amplitude and the classical double copies in the self-dual sector.
Finally, we briefly mention that the extension of this discussion to the higher-spin case suggests that scalar point particles cannot interact with chiral higher-spin fields.
 }

\begin{document} 
\maketitle
\flushbottom

\section{Introduction}

Color-kinematics duality \cite{Bern:2008qj,Bern:2010ue} is a statement that scattering amplitudes in gauge theories can be properly rearranged, so that the kinematic blocks they feature satisfy the same relations as the color factors. This structure hints towards the fact that in addition to the color algebra there exists its counterpart -- the kinematic algebra -- which plays a central role in the gauge theory perturbative dynamics, in particular, by constraining the kinematic dependence of scattering amplitudes. Besides that, the duality states that gauge theory scattering amplitudes give rise to gravity amplitudes via the double-copy procedure, which amounts to the replacement of the color factors in the gauge theory graph by another copy of the kinematic factors. Thus,  color-kinematics duality not only uncovers some form of a hidden symmetry underlying  gauge theory amplitudes, but it also reveals a deep and highly non-trivial connection between gauge theory and gravity. On the practical side, considering that the perturbative expansion of the gravitational action is by far more complex than the expansion of its gauge theory counterpart,
the duality proved to be particularly instrumental for perturbative computations in gravity at high loop orders \cite{Bern:2010tq,Bern:2012uc,Bern:2013uka}.

Since the duality was discovered it attracted considerable interest. It was found that its scope goes well beyond pure Yang-Mills and gravity theories. In particular, it applies to theories with supersymmetry, theories with various types of matter, to string theories, etc., connecting them into an intricate duality web. The duality was also extended away from the original setup of flat space scattering amplitudes. In particular,  classical color-kinematics duality, which relates non-perturbative classical solutions in gauge theory and gravity, was suggested \cite{Monteiro:2014cda,Luna:2015paa}. Despite sharing many common features with the amplitude version of the duality,  classical color-kinematics duality largely remains an independent construction, see, however, recent discussions in \cite{Kim:2019jwm,Monteiro:2020plf,Crawley:2021auj,Guevara:2021yud,Monteiro:2021ztt,Luna:2022dxo}.
Reviews on various aspects of color-kinematics duality as well as  more extensive references to the literature can be found in \cite{Carrasco:2015iwa,Bern:2019prr,Bern:2022wqg,Kosower:2022yvp,Adamo:2022dcm}.

More recently, an extension of  color-kinematics duality to point particles interacting with gauge theory and gravitational fields attracted considerable interest. It is motivated by the following two reasons. Firstly, regarding point particles as sources,  one can compute the associated classical solutions in gravity and gauge theory perturbatively \cite{Luna:2016hge,Monteiro:2020plf,Guevara:2020xjx}. This gives a setup, which is intermediate between that of scattering amplitudes and that of non-perturbative classical solutions. Thus, color-kinematics duality for point particles may eventually help bridging the gap between  amplitude and  classical color-kinematics dualities. Secondly, recent experimental successes in gravitational-wave physics stimulated complementary advances in theoretical modelling of binary sources and the associated gravitational-wave signals. Modern amplitude techniques proved to be very efficient in computing  classical observables, allowing to obtain state-of-the-art predictions relevant for classical scattering, see e. g. \cite{Bern:2019crd,Bern:2021yeh}. These observables are similar to amplitudes, the main difference being that in addition to quantum fields the former involve black holes, modelled by classical  point particles.
These can be accessed by starting from suitable amplitudes and taking the appropriate classical limit, see \cite{Kosower:2018adc,Maybee:2019jus,delaCruz:2020bbn,Cristofoli:2021vyo,Aoude:2021oqj}. Alternatively, the relevant observables can be computed more directly by using various expansion schemes featuring point particles from the very beginning \cite{Goldberger:2004jt,Neill:2013wsa,Porto:2016pyg,Mogull:2020sak,Jakobsen:2021zvh}. Irrespectively of the concrete setup, color-kinematics duality plays an important role in the analysis as it allows one to reduce more complex computations involving gravitational fields to simpler gauge theory ones.

Color-kinematics duality for point particles was studied  at the level of classical perturbative observables quite extensively, see
\cite{Goldberger:2016iau,Goldberger:2017frp,Luna:2017dtq,Goldberger:2017vcg,Goldberger:2017ogt,Chester:2017vcz,Li:2018qap,Shen:2018ebu,Plefka:2018dpa,Plefka:2019hmz,Bastianelli:2021rbt,Shi:2021qsb}. Besides that, in \cite{Gonzo:2021drq,Ball:2023xnr} it was found that  for point particles propagating in double-copy-related backgrounds the associated conserved quantities are related by some form of the double-copy procedure.

In the present paper we focus on  color-kinematics duality for point particles at the level of action. To this end we consider a special setup of self-dual fields.
 What makes self-dual field theories special is that in the light-cone gauge these involve only cubic vertices \cite{Plebanski:1975wn,Park:1990fp,Siegel:1992wd,Bardeen:1995gk,Cangemi:1996rx,Chalmers:1996rq}. Due to that, the Feynman rules immediately produce (off-shell) amplitudes in the trivalent form and no rearrangements are needed for  color-kinematics duality to be seen. Equivalently, this implies that the duality becomes manifest at the level of action. 
 This special structure of self-dual theories was utilised in \cite{Monteiro:2011pc} and, in particular, the kinematic algebra in the given case was identified. For the follow-up developments within this framework, see e.g. \cite{Boels:2013bi,Armstrong-Williams:2022apo,Monteiro:2022nqt,Doran:2023cmj}.

Below we extend this result to point particles interacting with a self-dual gauge field and gravity. We find that, within the appropriate light-cone Hamiltonian formulations of the worldline theories the associated actions become very simple and exhibit manifest color-kinematics duality.
A similar structure also holds for scalar fields interacting minimally with self-dual Yang-Mills theory and gravity. By applying a certain limiting procedure that replaces scalar fields with point particles, we are able to reproduce the worldline theories we considered initially.  We, therefore, conclude that color-kinematics duality for worldline theories interacting with self-dual fields is naturally inherited from the duality in the field theory case via the aforementioned limiting procedure.

This paper is organised as follows. In section \ref{sec2} we consider point particles  interacting with a self-dual Yang-Mills field and self-dual gravity. We start from the standard covariant point particle actions. By carrying out the Hamiltonian analysis with light-like time variable we arrive at the actions in the desired form. In section \ref{sec3} we show that these actions  exhibit manifest color-kinematics duality. Next, in section \ref{sec4} we consider scalar fields interacting minimally with a self-dual Yang-Mills field and self-dual gravity. We then discuss color-kinematics duality in this case. 
In section \ref{sec5} we show that these scalar  field theories  result in the point particle theories  we started from within the appropriate limit. In this way, we relate color-kinematics dualities in the two setups. Finally, we conclude in section \ref{sec6}.

\section{Point particle actions}
\label{sec2}
Color-kinematics duality for point particles becomes manifest only once we use the  Hamiltonian formalism with the appropriately chosen time. It is also required that the  Yang-Mills field and gravity are self-dual and the light-cone gauge is imposed. In this section we will start with the familiar covariant actions and carry out the necessary manipulations bringing them to the desired form.

\subsection{Point particle coupled to a self-dual Yang-Mills field}

A point particle can be coupled to  a Yang-Mills  field in the following way \cite{Balachandran:1976ya,Sternberg:1977rf}\footnote{A different action can be found in \cite{Balachandran:1977ub}.}
\begin{equation}
\label{19feb1}
S=-m\int d\tau \sqrt{\eta_{\mu\nu}\frac{d x^\mu}{d\tau} \frac{d x^\nu}{d\tau} }+\int d\tau \left(i \theta_a^\dagger \frac{d\theta_a}{d\tau} - g \frac{d x^\mu}{d\tau}  A_\mu^\alpha \theta_a^\dagger T^\alpha_{ab} \theta_b\right),
\end{equation}
where $\theta$ and $\theta^\dagger$ represent additional point particle color degrees of freedom transforming in some representation of the color symmetry algebra and its dual respectively. We will not distinguish  upper and lower color indices assuming that these can be raised or lowered by the unit matrix.

A self-dual Yang-Mills field can be expressed in the light-cone gauge as follows \cite{Bardeen:1995gk,Cangemi:1996rx,Chalmers:1996rq,Monteiro:2011pc}
\begin{equation}
\label{19feb2}
A^\alpha_v = -\frac{1}{4}\partial_w \Phi^\alpha, \qquad A^\alpha_{\bar w}=-\frac{1}{4}\partial_u \Phi^\alpha
\end{equation}
with the other components vanishing.
Here we use the light-cone coordinates
\begin{equation}
\label{19deb3}
u = t-z, \qquad v = t+z, \qquad w = x+iy, \qquad \bar w = x-iy.
\end{equation}
In these coordinates the metric is defined by
\begin{equation}
\label{19feb4}
ds^2 = du dv - dw d\bar w.
\end{equation}

As the next step, we fix the reparametrization invariance of (\ref{19feb1}) by setting $\tau = v$. This leads to
\begin{equation}
\label{19feb5}
S=-m\int dv \sqrt{\dot u - \dot w \dot{\bar w}} + 
\int dv \left(i \theta_a^\dagger \dot\theta_a +\frac{g}{4}(\partial_w \Phi^\alpha + \dot{\bar w} \partial_u \Phi^\alpha)\theta_a^\dagger T_{ab}^\alpha \theta_b \right),
\end{equation} 
where we use dots to denote derivatives with respect to $v$.

Next, we move on to the Hamiltonian formalism. The canonical spatial momenta are
\begin{equation}
\label{19feb6}
\begin{split}
p_u & \equiv \frac{\partial L}{\partial \dot u}=-\frac{m}{2\sqrt{\dot u -\dot w \dot{\bar w}}}, \\
p_w  &\equiv \frac{\partial L}{\partial \dot w} = \frac{m \dot{\bar w}}{2 \sqrt{\dot u -\dot w \dot{\bar w}}},\\
p_{\bar w} & \equiv \frac{\partial L}{\partial \dot{\bar w}}= 
 \frac{m \dot{ w}}{2 \sqrt{\dot u -\dot w \dot{\bar w}}}+\frac{g}{4} \partial_u \Phi^\alpha \theta_a^\dagger T_{ab}^\alpha \theta_b
\end{split}
\end{equation}
and the canonical commutation relations are given by
\begin{equation}
\label{19feb6x1}
\{x^i,p_j \}= \delta^i{}_j, \qquad i,j = u,w,\bar w.
\end{equation}
Equations (\ref{19feb6}) can be solved for velocities, which leads to
\begin{equation}
\label{19feb7}
\begin{split}
\dot u &= \frac{m^2}{4p^2_u}+ \frac{p_w}{p_u}\frac{p_{\bar w}-\frac{g}{4} \partial_u \Phi^\alpha \theta_a^\dagger T_{ab}^\alpha \theta_b}{p_u},\\ 
\dot w &= -\frac{p_{\bar w}-\frac{g}{4}\theta_a^\dagger T_{ab}^\alpha \theta_b}{p_u},\\
\dot{\bar w} &=-\frac{p_w}{p_u}.
\end{split}
\end{equation}

Color coordinates $\theta_a$ and $\theta_a^\dagger$ are regarded as independent. Accordingly, we have the following color momenta
\begin{equation}
\label{19feb8}
\begin{split}
p_{\theta_a} &\equiv \frac{\partial L}{\partial \dot{\theta_a}} = i \theta^{\dagger}_a, \\
p_{\theta^\dagger_a} &\equiv \frac{\partial L}{\partial \dot{\theta^\dagger_a}} =0.
\end{split}
\end{equation}
The associated Poisson brackets read
\begin{equation}
\label{19feb8x1}
\{\theta_a, p_{\theta_b} \} = \delta_{ab}, \qquad \{\theta_a^\dagger, p_{\theta^\dagger_b} \}=\delta_{ab}.
\end{equation}

Both  equations (\ref{19feb8}) present constraints
\begin{equation}
\label{19feb9}
C_a\equiv p_{\theta_a}-i\theta^\dagger_a =0,\qquad C^\dagger_a \equiv p_{\theta^\dagger_a}=0.
\end{equation}
It is not hard to see \cite{Balachandran:1976ya} that $\dot{C_a} =0$ and $\dot{C^\dagger_a}=0$ allow one to solve for $C_a$ and $C^\dagger_a$, thus, the system does not have secondary constraints. Moreover, $C_a$ and $C^\dagger_a$ do not commute, thus presenting a pair of second class constraints. Eliminating $p_{\theta_a}$ and $p_{\theta^\dagger_a}$, we find a reduced phase space with the only color coordinates $\theta_a$ and $\theta^\dagger_a$. The associated Dirac bracket reads
\begin{equation}
\label{19feb10}
\{\theta_a,\theta_b^\dagger \}_D=-i\delta_{ab}.
\end{equation}

By definition, the Hamiltonian is
\begin{equation}
\label{19feb11}
\begin{split}
H&\equiv p_u \dot u + p_w\dot w+ p_{\dot{\bar w}}\dot{\bar w} + p_{\theta_a} \dot{\theta_a}+p_{\theta_a^\dagger} \dot\theta_a^\dagger -L\\
&\qquad\qquad\qquad = - \frac{m^2}{4p_u}- \frac{p_w p_{\bar w}}{p_u}-\frac{g}{4}\Big(\partial_w - \frac{p_w}{p_u}\partial_u \Big) \Phi^\alpha 
 \theta_a^\dagger T_{ab}^\alpha \theta_b.
\end{split}
\end{equation}
Accordingly, the action can be written as 
\begin{equation}
\label{20feb2}
S= S_0 + S_{int},
\end{equation}
where
\begin{equation}
\label{20feb1}
\begin{split}
S_0 &=\int dv \left(p_u \dot u + p_w\dot w+ p_{\dot{\bar w}}\dot{\bar w} + p_{\theta_a} \dot{\theta_a}+p_{\theta_a^\dagger} \dot\theta_a^\dagger 
+ \frac{m^2}{4p_u}+ \frac{p_w p_{\bar w}}{p_u}\right),\\
S_{int}&=-\int dv H_{int} = \frac{g}{4}\int \frac{dv}{p_u} \left(p_u\partial_w - {p_w}\partial_u \right)  \theta_a^\dagger T_{ab}^\alpha \theta_b \Phi^\alpha .
 \end{split}
\end{equation}

\subsection{Point particle coupled to a self-dual gravitational field}

In a similar manner we treat a point particle interacting minimally with a self-dual gravitational field. The initial covariant action is
\begin{equation}
\label{19feb12}
S= - m \int d\tau \sqrt{g_{\mu\nu}\frac{d x^\mu}{d\tau}\frac{d x^\nu}{d\tau}}.
\end{equation}
A self-dual metric can be presented in the form \cite{Plebanski:1975wn,Monteiro:2011pc}
\begin{equation}
\label{19feb13}
g_{\mu\nu} = \eta_{\mu\nu}+\kappa h_{\mu\nu},
\end{equation}
where
\begin{equation}
\label{19feb14}
h_{vv}=-\frac{1}{4}\partial^2_{w}\Phi, \qquad h_{\bar w\bar w}=-\frac{1}{4} \partial^2_u \Phi, \qquad h_{v\bar w}= h_{\bar w v}= -\frac{1}{4} \partial_w \partial_u \Phi
\end{equation}
and other components are vanishing.
Fixing the reparametrization invariance as $\tau =v$ and using (\ref{19feb14}), we find
\begin{equation}
\label{19feb15}
S=-m \int dv \sqrt{\dot u - \dot w \dot{\bar w}-\frac{\kappa}{4}(\partial_w^2 \Phi +\partial_u^2 \Phi (\dot{\bar w})^2 +2 \partial_w\partial_u \Phi \dot{\bar w})}.
\end{equation}

The canonical momenta are
\begin{equation}
\label{19feb16}
\begin{split}
p_u& \equiv \frac{\partial L}{\partial \dot u}=-\frac{m}{2\sqrt{\dot u -\dot w \dot{\bar w} - \frac{\kappa}{4} (\partial_w + \dot{\bar w}\partial_u)^2\Phi}}, \\
p_w &\equiv \frac{\partial L}{\partial \dot w} = \frac{m \dot{\bar w}}{2\sqrt{\dot u -\dot w \dot{\bar w} - \frac{\kappa}{4} (\partial_w + \dot{\bar w}\partial_u)^2\Phi}},\\
p_{\bar w}& \equiv \frac{\partial L}{\partial \dot{\bar w}}=  \frac{m \dot{w} +\frac{\kappa}{2}(\dot{\bar w}\partial^2_u \Phi +\partial_u\partial_w \Phi)}{2\sqrt{\dot u -\dot w \dot{\bar w} - \frac{\kappa}{4} (\partial_w + \dot{\bar w}\partial_u)^2\Phi}}.
\end{split}
\end{equation}
The Poisson bracket is the same as in (\ref{19feb6x1}).
Equations (\ref{19feb16}) can be solved for velocities, which leads to
\begin{equation}
\label{19feb20}
\begin{split}
\dot u & = \frac{m^2}{4p_u^2}- \frac{p_w}{p_u}\left(-\frac{p_{\bar w}}{p_u} +\frac{\kappa}{2} \left( \frac{p_w}{p_u}\partial_u - \partial_w \right)\partial_u \Phi \right) +\frac{\kappa}{4}\left(\partial_w - \frac{p_w}{p_u}\partial_u \right)^2\Phi,\\
\dot w & = - \frac{p_{\bar w}}{p_u}+\frac{\kappa}{2} \left(\frac{p_w}{p_u}\partial_u - \partial_w \right)\partial_u\Phi,\\
\dot{\bar w} & = -\frac{p_w}{p_u}.
\end{split}
\end{equation}

Eventually, we find the Hamiltonian
\begin{equation}
\label{19feb21}
\begin{split}
H\equiv p_u \dot u + p_w\dot w+ p_{\dot{\bar w}}\dot{\bar w}  -L
= - \frac{m^2}{4p_u}- \frac{p_w p_{\bar w}}{p_u}+\frac{\kappa}{4}p_u \left(\partial_w - \frac{p_w}{p_u}\partial_u \right)^2 \Phi.
\end{split}
\end{equation}
Accordingly, the action split into the free and the interacting parts, (\ref{20feb2}), reads
\begin{equation}
\label{20feb3}
\begin{split}
S_0 &= \int dv \left(p_u \dot u + p_w\dot w+ p_{\dot{\bar w}}\dot{\bar w}  
+ \frac{m^2}{4p_u}+ \frac{p_w p_{\bar w}}{p_u}\right),\\
S_{int} &=-\int dv H_{int} =-\frac{\kappa}{4}\int  \frac{dv}{p_u} \left(p_u\partial_w - {p_w}\partial_u \right)^2 \Phi.
\end{split}
\end{equation}

Let us note that, quite remarkably, despite the initial covariant action (\ref{19feb12}) features a square root, the light-cone Hamiltonian action (\ref{20feb3}) only involves a single interaction vertex, which is linear in fluctuations of the gravitational field. This vertex is very similar in structure to its Yang-Mills counterpart (\ref{20feb1}). As we will now show, this similarity is a manifestation of color-kinematics duality.

\section{Color-kinematics duality}
\label{sec3}

In the present section we will consider Hamiltonians (\ref{20feb1}), (\ref{20feb3}) and the associated actions in the context of color-kinematics duality.
 As we have found, these actions feature only a single interaction term in each case, and these are linear in fields. This structure mimics closely that of self-dual Yang-Mills theory and self-dual gravity \cite{Monteiro:2011pc}. As in the latter case, we expect that this makes color-kinematics duality  manifest. 

We first consider the vertex (\ref{20feb1}) for the self-dual Yang-Mills case. With $(p_u)^{-1}$ understood as a measure, it factorises into
\begin{equation}
\label{20feb4}
c\equiv \theta_a^\dagger T_{ab}^\alpha \theta_b  \qquad \text{and} \qquad n\equiv p_u\partial_w - {p_w}\partial_u,
\end{equation}
which we regard as the color and the kinematic factors respectively.

Let us first focus on the color factor. It is straightforward to compute that
\begin{equation}
\label{20feb5}
\big{\{} \theta_a^\dagger T_{ab}^\alpha \theta_b \varepsilon_1(x), \theta_a^\dagger T_{ab}^\alpha \theta_b \varepsilon_2(x) \big{\}}_D =
-i \theta_a^\dagger [T^\alpha,T^\beta]_{ab}\theta_b \varepsilon_1(x) \varepsilon_2(x),
\end{equation}
where $\varepsilon_i(x)$ are arbitrary functions, to be understood as the color algebra gauge pa\-ra\-me\-ters. It is straightforward to see that (\ref{20feb5}) implies that the  Dirac bracket
\begin{equation}
\label{20feb6}
\{\theta_a^\dagger T_{ab}^\alpha \theta_b \varepsilon(x), \quad \cdot \quad   \}_D
\end{equation}
realises a representation of the color algebra defined by $T$ on point particle's phase space. 

In a similar manner, we consider the kinematic factor. One can show that
\begin{equation}
\label{20feb7}
\begin{split}
\{ (p_u \partial_w - p_w \partial_u)\varepsilon_1(x), (p_u \partial_w - p_w \partial_u)\varepsilon_2(x) \} 
=(p_w \partial_w - p_w\partial_u)[\varepsilon_1,\varepsilon_2](x),
\end{split}
\end{equation}
where 
\begin{equation}
\label{20feb8}
[\varepsilon_1,\varepsilon_2](x)= \partial_u \varepsilon_1(x) \partial_w \varepsilon_2 (x) - \partial_w \varepsilon_1(x) \partial_u \varepsilon_2(x)
\end{equation}
is the Lie commutator of the algebra of area-preserving diffeomorphisms. 
This algebra was identified in \cite{Monteiro:2011pc} as the kinematic algebra of self-dual Yang-Mills theory.
Here again $\varepsilon_i(x)$ are arbitrary functions which are understood as gauge parameters of the kinematic algebra. 
 Equation (\ref{20feb7}) implies that  the Poisson bracket
\begin{equation}
\label{20feb9}
\{(p_u \partial_w - p_w \partial_u)\varepsilon (x), \quad \cdot \quad   \} 
\end{equation}
realises a representation of this algebra on point particle's phase space\footnote{It would be interesting to explore a potential role of this conclusion in celestial holography, see e.g. \cite{Campiglia:2021srh,Strominger:2021mtt,Adamo:2021lrv,Monteiro:2022lwm} for recent discussions of self-dual theories in the holographic context.}. 

Identities (\ref{20feb5}) and (\ref{20feb7}) can be interpreted as the worldline counterpart of the BCJ relations. Indeed, the BCJ relations state that the kinematic factors satisfy the same identities as the color ones. This is exactly what (\ref{20feb5}) and (\ref{20feb7}) state in the point-particle case.
More precisely, they imply that the kinematic and the color factors define generators of the kinematic and the color algebra respectively on point particle's phase space.

Finally, we note that the gravitational vertex (\ref{20feb4}) quite manifestly has the double-copy structure: with the $(p_u)^{-1}$ factor absorbed into the measure, it has the form of the kinematic factor squared. Thus, both features required by color-kinematic duality -- the BCJ relations and the double copy -- are present for worldline actions (\ref{20feb1}), (\ref{20feb3}).

Considering that the structure of the worldline actions we found here is very reminiscent of that for self-dual Yang-Mills theory and gravity \cite{Monteiro:2011pc}, it seems reasonable to try to establish the connection between the two setups.
It is more natural, however, to compare the point particle theories we are dealing with not with  self-dual Yang-Mills theory and self-dual gravity, but rather with the theories of scalar fields interacting with a self-dual Yang-Mills field and self-dual gravity. This will be done in the following sections. This analysis will also explain the origin of $(p_u)^{-1}$ measure factors that we  encountered above.

\section{Field theory actions and color-kinematics duality}
\label{sec4}

In the present section we will consider  a scalar field minimally coupled to a self-dual Yang-Mills field and to self-dual gravity. To make contact with the previous discussion, these theories will be written in the light-cone coordinates and with the light-cone gauge imposed.

We start from the standard covariant action in the Yang-Mills case
\begin{equation}
\label{20feb10}
S=\frac{1}{2}\int d^4 x \left(D^\mu \phi^a D_\mu \phi^a - m^2 \phi^a \phi^a \right), \qquad D_\mu \phi^a = \partial_\mu \phi^a +ig A_\mu^\alpha T_{ab}^\alpha \phi^b.
\end{equation}
We may ignore the $\sqrt{-\eta}$ factor, as it is constant.
By using the light-cone metric (\ref{19feb4}) and representation (\ref{19feb2}) for the gauge field, we arrive at
\begin{equation}
\label{20feb11}
\begin{split}
&S=\frac{1}{2}\int d^4x \left(4\partial_u \phi^a \partial_v\phi^a - 4 \partial_w \phi^a \partial_{\bar w}\phi^a-m^2 \phi^a\phi_a\right)\\
 &\qquad\qquad\qquad\qquad\qquad-
i\frac{g}{2} \int d^4x \left(\partial_u \phi^a \partial_w \Phi^\alpha - \partial_w \phi^a \partial_u \Phi^\alpha \right) T^{\alpha}_{ab}\phi^b.
\end{split}
\end{equation}
Using the Bose symmetry and integration by parts, one can see that the symmetric part of $T_{ab}$ drops out. Thereby, without loss of generality, we can assume $T^\alpha_{ab}=-T^\alpha_{ba}$.

In a similar manner, we consider the covariant action for the gravity case
\begin{equation}
\label{20feb12}
S=\frac{1}{2}\int d^4x(g^{\mu\nu}\partial_\mu \phi \partial_\nu \phi -m^2 \phi^2).
\end{equation}
For  metric given in the form (\ref{19feb13}), (\ref{19feb14}), the inverse metric reads
\begin{equation}
\label{20feb13}
\begin{split}
g^{uv}&=g^{vu}=2, \qquad g^{w\bar w}=g^{\bar w w}=-2, \\
 g^{uu} &=\kappa \partial^2_w \Phi, \qquad g^{ww}= \kappa\partial^2_u\Phi, \qquad g^{uw}=g^{wu}=-\kappa\partial_w\partial_u \Phi
 \end{split}
\end{equation}
with the remaining components vanishing.
Considering that $\sqrt{-g}=\frac{1}{4}$ is just a constant, we dropped this factor in (\ref{20feb12}).
Employing (\ref{20feb13}), we arrive at the light-cone form of the action
\begin{equation}
\label{20feb14}
\begin{split}
S&=\frac{1}{2} \int d^4x (4\partial_u \phi \partial_v \phi - 4 \partial_w \phi \partial_{\bar w}\phi - m^2 \phi^2)\\
&\qquad \qquad \qquad \qquad -\frac{\kappa}{2} \int d^4x 
 \left(\partial^2_w \Phi \partial^2_u\phi  - 2 \partial_w \partial_u \Phi \partial_u \partial_w\phi +\partial^2_u \Phi \partial^2_w \phi\right) \phi.
\end{split}
\end{equation}

Quite remarkably, in both cases (\ref{20feb11}), (\ref{20feb14}) the action truncates at the cubic order. Moreover, it is not hard to see that these actions satisfy color-kinematics duality at the level of action. To show this, we follow the steps from \cite{Monteiro:2011pc}\footnote{Strictly speaking, \cite{Monteiro:2011pc} only deals with equations of motions for self-dual theories. This discussion can be easily promoted to the level of action, which, however, requires adding fields of opposite helicities.}. 

To this end, we first go to the momentum space. Then, up to a constant factor, the cubic vertex (\ref{20feb11}) has a kernel
\begin{equation}
\label{7mar1}
\delta^4(p_1+k-p_2)(p_{1w}k_{u}-p_{1u}k_{w})T^{\alpha}_{ab},
\end{equation}
where $k$ refers to the momentum of the gauge field, $p_1$ refers to the momentum of one of the scalar fields, while $-p_2$ is the momentum of the remaining scalar field. The reason why the momentum of the second scalar field comes with the minus sign is that the natural measure in momentum space is
\begin{equation}
\label{7mar2}
\langle \phi_1|\phi_2\rangle =\int d^4p \phi_1(p)\phi_2(-p)
\end{equation}
and, accordingly, by changing the sign of momentum one raises a ''momentum index''.
Putting it differently, instead of a convention in which all momenta are ingoing, in (\ref{7mar1}) one uses a convention in which one scalar field is ingoing, while the other one is outgoing. The latter convention is more appropriate, since we want to interpret the kernel (\ref{7mar1}) in terms of transformations acting on scalar fields.

From the kernel (\ref{7mar1}) one extracts the color and the kinematic factors
\begin{equation}
\label{7mar3}
C\equiv \delta^4(p_1+k-p_2)T^{\alpha}_{ab}, \qquad N \equiv \delta^4(p_1+k-p_2)(p_{1w}k_{u}-p_{1u}k_{w}).
\end{equation}
These define transformations of the scalar field 
\begin{equation}
\label{18mar1}
\begin{split}
(\delta^C_\varepsilon \phi)^a (p_2) &=\int d^4 p_1d^4 k \delta^4(p_1+k-p_2)T^{\alpha}_{ab} \phi^b(p_1)\varepsilon^\alpha(k),\\
(\delta^N_\varepsilon \phi) (p_2)&=\int d^4 p_1d^4 k \delta^4(p_1+k-p_2)(p_{1w}k_{u}-p_{1u}k_{w}) \phi(p_1)\varepsilon(k)
\end{split}
\end{equation}
in the color and the kinematic cases respectively. Here $\varepsilon$ are again understood as parameters of symmetry transformations.
It is not hard to see that (\ref{18mar1}) define representations of the color and the kinematic algebras realised on the space of states of the scalar field. Finally, by squaring the kinematic factor and dropping a redundant momentum conservation, we obtain the kernel of the vertex in the gravitational case (\ref{20feb14}).

This discussion is a straightforward generalisation of the one given in \cite{Monteiro:2011pc} for self-dual Yang-Mills theory and for self-dual gravity to scalar fields interacting minimally with a self-dual Yang-Mills field and self-dual gravity. At the same time, we are not aware of it having been presented previously, see \cite{Chiodaroli:2013upa,Johansson:2014zca,Chiodaroli:2015rdg,Bern:2019prr} for previous literature on color-kinematics duality for theories with scalar matter\footnote{The fact that actions (\ref{20feb11}), (\ref{20feb14}) are Lorentz invariant follows from general arguments of \cite{Ponomarev:2017nrr}. Still, their connection to covariant actions (\ref{20feb10}), (\ref{20feb12}) is not obvious without a simple computation we carried out above.}.

\section{Limiting procedure}
\label{sec5}

In this section, we will present a limiting procedure, that relates field-theory  (\ref{20feb11}), (\ref{20feb14}) and  point-particle actions 
(\ref{20feb1}), (\ref{20feb3}).

In principle, to connect classical field-theory and point-particle actions directly,  one can imagine a procedure which would amount to the substitution of localised field configurations -- designed to mimic world lines of point particles -- into the action of a field theory, thus, getting a point-particle action. It does not seem, however, that this type of a procedure has a solid physical basis. Instead, we will consider a more complex construction, which is analogous to  \cite{Kosower:2018adc,delaCruz:2020bbn}, except that it works at the level of the Hamiltonian and not at the level of scattering observables.

The limit that we are going to employ consists of few steps.  As the first step we are going to quantise the scalar field of the classical field theory. This step is what introduces a concept of a particle into the discussion. Considering that the actions we are dealing with are quadratic in scalar fields, the associated quantum theories  do not involve particle creation or annihilation.
As a result, one-particle states form a closed sector of the Hilbert space of these theories. Restricting ourselves to this sector we end up with a theory of relativistic quantum mechanics.  Finally, considering the classical limit, we will arrive at a classical point-particle theory. We are going to carry out these steps at the level of the Hamiltonian, as it has the universal meaning of the evolution operator irrespectively of whether the theory is classical or quantum and whether we are dealing with fields or point particles.

\subsection{Gravity case}

We start from the self-dual gravity case, as it is more straightforward. As we have just explained, first, we will quantise the scalar field in (\ref{20feb14}). We will do that using the Hamiltonian quantisation.

Still regarding $v$ as the time variable, we compute the canonical momenta
\begin{equation}
\label{20feb15}
\pi(x) \equiv \frac{\delta L}{\delta \dot \phi}(x) = 2\partial_u\phi(x).
\end{equation}
The associated Poisson bracket is
\begin{equation}
\label{20feb16}
\{ \phi(x),\partial_u\phi(y)\} =\frac{1}{2}\delta^3 (x-y).
\end{equation}
Computing the Hamiltonian, we find
\begin{equation}
\label{20feb17}
\begin{split}
H&\equiv \int d^3 x (\pi(x) \partial_v \phi(x) - L(x)) =2\int d^3 x (\partial_w \phi \partial_{\bar w}\phi + \frac{m^2}{4}\phi^2) \\
& \qquad \qquad\qquad \qquad
+\frac{\kappa}{2}\int d^3 x\left(\partial^2_w \Phi \partial^2_u\phi  - 2 \partial_w \partial_u \Phi \partial_u \partial_w\phi +\partial^2_u \Phi \partial^2_w \phi \right) \phi.
\end{split}
\end{equation}

Equation (\ref{20feb15}) gives a constraint. Its treatment is well-known, see e.g. \cite{Perry:1994kp}, and will not be repeated here. Eventually, the Poisson bracket should be replaced with the Dirac bracket which is just 
\begin{equation}
\label{20feb18}
\{ \phi(x),\partial_u\phi(y)\}_D =\frac{1}{4}\delta^3 (x-y).
\end{equation}

When quantising a theory, the Dirac bracket should be deformed into the quantum commutator
\begin{equation}
\label{20feb19}
[\hat\phi(x),\partial_u\hat\phi(y)] =\frac{i}{4}\delta^3 (x-y).
\end{equation}
By replacing $\phi$'s with their operator versions in (\ref{20feb17}), we obtain a quantum field theory Hamiltonian $\hat H$.

The standard quantization procedure then requires to find a representation of (\ref{20feb19}) in terms of creation and annihilation operators. Such a representation is well-known, see e.g. \cite{Perry:1994kp}. We will not give it here, because, as we will see shortly, for our purposes it suffices to know the commutator of fields (\ref{20feb19}), while the detailed realisation of fields in terms of creation and annihilation operators is not needed.

Our next step is to understand how $\hat H$ acts on single-particle states
\begin{equation}
\label{20feb20}
| \phi(x)\rangle \equiv \hat\phi(x)|0\rangle,
\end{equation}
where $|0\rangle$ is the vacuum. To evaluate the action of the Hamiltonian on $| \phi\rangle$, we proceed as
\begin{equation}
\label{20feb21}
\hat H \hat\phi(x) |0\rangle = [\hat H, \hat\phi (x)] |0\rangle +  \hat\phi(x)  \hat H |0\rangle.
\end{equation}
The last term on the right-hand side can be dropped, as usual, by arguing that the definition of energy can be always shifted, so that  for the vacuum state the energy is vanishing. By evaluating the remaining term in (\ref{20feb21}) by means of (\ref{20feb19}), we find
\begin{equation}
\label{20feb22}
\begin{split}
\hat H |\phi(x)\rangle& = -i \left( \frac{\partial_w\partial_{\bar w}}{\partial_u} - \frac{m^2}{4}\frac{1}{\partial_u} \right)|\phi(x)\rangle 
\\& \qquad\qquad\qquad-
\frac{\kappa}{4} \frac{1}{\partial_u} \left(\partial^2_w \Phi \partial^2_u  - 2\partial_w \partial_u\Phi \partial_u\partial_w +\partial_u^2 \Phi \partial_w^2 \right)|\phi(x)\rangle,
\end{split}
\end{equation}
where in the second line, it is assumed that $\partial_u^{-1}$ acts on both $\Phi$ and $|\phi\rangle$. Formula (\ref{20feb22}) defines the action of the Hamiltonian operator on the relativistic quantum mechanical wave function.

Finally, we consider the classical limit of (\ref{20feb22}). This amounts to the replacement $\partial_j \to - ip_j$ for derivatives acting on the wave function. Besides that, in the classical limit $p$ is assumed to be large, so $(\partial_u)^{-1}$ in the second line of (\ref{20feb22}) may be considered as acting on $|\phi\rangle$ only. Eventually, we find the classical worldline Hamiltonian  (\ref{19feb21}).

\subsection{Yang-Mills case}

Similarly, we treat the Yang-Mills case. For action (\ref{20feb11}) the canonical momenta are
\begin{equation}
\label{21feb1}
\pi^a(x) \equiv \frac{\delta L}{\delta \dot \phi^a}(x) = 2\partial_u\phi^a(x).
\end{equation}
The Hamiltonian then reads
\begin{equation}
\label{21feb2}
\begin{split}
H&\equiv \int d^3 x (\pi^a(x) \partial_v \phi^a(x) - L(x)) \\
&=2\int d^3 x (\partial_w \phi^a \partial_{\bar w}\phi^a + \frac{m^2}{4}\phi^a\phi^a)
+i\frac{g}{2}\int d^3 x\left(\partial_w \Phi^\alpha \partial_u\phi^a -\partial_u \Phi^\alpha \partial_w \phi^a \right)T^\alpha_{ab} \phi^b.
\end{split}
\end{equation}
As in the gravity case, equations (\ref{21feb1}) present constraints, which are treated the same way. Eventually, after quantisation, we find the bracket
\begin{equation}
\label{21feb3}
[\hat\phi^a(x),\partial_u\hat\phi^b(y)] =\frac{i}{4}\delta^3 (x-y)\delta^{ab}.
\end{equation}
Proceeding to the action of the Hamiltonian on one-particle states, we find
\begin{equation}
\label{21feb4}
\begin{split}
\hat H |\phi^a(x)\rangle &= [\hat H, \hat\phi^a(x)] |0\rangle \\
&= -i \left( \frac{\partial_w\partial_{\bar w}}{\partial_u} - \frac{m^2}{4}\frac{1}{\partial_u} \right)|\phi^a(x)\rangle
+\frac{g}{4}\frac{1}{\partial_u}(\partial_w \Phi^\alpha \partial_u - \partial_u \Phi^\alpha \partial_w)T^{\alpha}_{ba}|\phi^b(x)\rangle,
\end{split}
\end{equation}
where we used antisymmetry $T^\alpha_{ab}=-T^\alpha_{ba}$.

It is convenient to group different components of the wave function into a single wave function employing an auxiliary color variable
\begin{equation}
\label{21feb5}
|\phi(x,\theta)\rangle \equiv | \phi^a(x)\rangle \theta^a.
\end{equation}
Then, the Hamiltonian (\ref{21feb4}) can be rewritten as
\begin{equation}
\label{21feb6}
\hat H = -i \left( \frac{\partial_w\partial_{\bar w}}{\partial_u} - \frac{m^2}{4}\frac{1}{\partial_u} \right) +
\frac{g}{4}\frac{1}{\partial_u}(\partial_w \Phi^\alpha \partial_u - \partial_u \Phi^\alpha \partial_w)T^{\alpha}_{ba} \theta^a \frac{\partial}{\partial \theta^b}.
\end{equation}
In the classical limit we have $\partial_j \to - ip_j$ and $\frac{\partial}{\partial \theta^a} \to -\theta^\dagger_a$, which allows us to reproduce the classical worldline Hamiltonian (\ref{19feb11}) from the earlier discussion.

We would like to remark that the step (\ref{21feb5}) was somewhat heuristic. By this we mean that the most natural quantization procedure for (\ref{19feb1}) leads to a wave function 
$|\phi(x,\theta)\rangle$ with arbitrary dependence on $\theta$, not only linear, as we had in (\ref{21feb5}). More details on this issue as well as on other quantization schemes can be found in \cite{Balachandran:1976ya,Sternberg:1977rf,Balachandran:1977ub}.

In summary, as we have just demonstrated, point-particle theories (\ref{20feb1}), (\ref{20feb3}) can be obtained by the appropriate limiting procedures from the field-theory counterparts (\ref{20feb11}), (\ref{20feb14}). In this limit the color and the kinematic factors of field theory (\ref{7mar3}) go over to the color and the kinematic factors of point particles (\ref{20feb4}).
This analysis also allows us to  connect the $(p_u)^{-1}$ measure factors in point-particle actions to the $\partial_u$ factor in the field theory Dirac bracket (\ref{20feb19}), (\ref{21feb3}).

\section{Conclusions}
\label{sec6}
We considered point particles interacting with self-dual Yang-Mills  and self-dual gravity fields.  We showed that once rewritten in the Hamiltonian form with a  light-like coordinate as the time variable, these theories exhibit manifest color-kinematics duality at the action level. This result generalises the analogous one from \cite{Monteiro:2011pc} to point particles. 

To strengthen the connection with the field theory case further, we considered scalar fields interacting minimally with self-dual Yang-Mills and self-dual gravity fields. These theories can also be rewritten in a way that color-kinematics duality becomes manifest at the action level. By considering the appropriate limiting procedure these field theories result in the worldline theories we started from. In this  limit the color and the kinematic factors of field theories go over into the color and the kinematic factors of particle theories. This allows us to demonstrate that color-kinematics duality for point particles is naturally inherited from the duality in the field theory case. 

We would like to emphasise that this argument connects  worldline color-kinematics duality to the field theory duality of the amplitude type. At the same time, by regarding point particles as sources, one can study how their color-kinematics duality manifests itself at the level of the associated field theory classical solutions. This analysis may prove to be helpful for establishing a precise connection between  amplitude and  classical color-kinematics dualities at least in the self-dual sector, see 
\cite{Kim:2019jwm,Monteiro:2020plf,Crawley:2021auj,Guevara:2021yud,Monteiro:2021ztt,Luna:2022dxo} for earlier works.
Besides that, the connection between the dualities in the two setups
 suggests that, similarly to the field-theory case, color-kinematics duality for point particles cannot be made manifest at the level of action beyond the self-dual sector, instead, it can only be seen at the level of scattering observables. 

As a final remark, we would like to mention implications of our work for the problem of interactions of point particles  with chiral higher-spin fields  \cite{Ivanovskiy:2023aay}. Chiral higher-spin theories \cite{Metsaev:1991mt,Ponomarev:2016lrm} can be regarded as natural higher-spin counterparts of self-dual gravity and self-dual Yang-Mills theories \cite{Ponomarev:2017nrr,Krasnov:2021nsq}, moreover, properly extended color-kinematics duality applies to chiral higher-spin theories as well  \cite{Ponomarev:2017nrr,Monteiro:2022xwq}. It is, therefore, natural to expect that the structures found in the present paper should also apply to the higher-spin case.
Among these, we would like to highlight the property that point-particle's phase space carries representations of both the color and the kinematic algebras. This property, of course, is not totally unexpected, as it can be regarded as the counterpart of the standard requirement that asymptotic states realise unitary irreducible representation of the global symmetry group in quantum field theory. At the same time, considering that the color and the kinematic algebras have parameters with functional freedom, the requirement that these should admit representations on point-particle's phase space can be very constraining. In particular, one can show that this argument rules out interactions of scalar point particles with chiral higher-spin fields. The details of this argument as well as the comprehensive analysis of interactions of point particles with chiral higher-spin fields will be given elsewhere.

\acknowledgments

We would like to thank A. Ochirov, P. Pichini and E. Skvortsov  for interesting discussions and for pointing out some relevant references. We would also like to thank  A. Ochirov and E. Skvortsov for comments on the manuscript.

\bibliography{pp}

\providecommand{\href}[2]{#2}\begingroup\raggedright\begin{thebibliography}{10}

\bibitem{Bern:2008qj}
Z.~Bern, J.J.M.~Carrasco and H.~Johansson, \emph{{New Relations for
  Gauge-Theory Amplitudes}},
  \href{https://doi.org/10.1103/PhysRevD.78.085011}{\emph{Phys. Rev. D}
  {\bfseries 78} (2008) 085011}
  [\href{https://arxiv.org/abs/0805.3993}{{\ttfamily 0805.3993}}].

\bibitem{Bern:2010ue}
Z.~Bern, J.J.M.~Carrasco and H.~Johansson, \emph{{Perturbative Quantum Gravity
  as a Double Copy of Gauge Theory}},
  \href{https://doi.org/10.1103/PhysRevLett.105.061602}{\emph{Phys. Rev. Lett.}
  {\bfseries 105} (2010) 061602}
  [\href{https://arxiv.org/abs/1004.0476}{{\ttfamily 1004.0476}}].

\bibitem{Bern:2010tq}
Z.~Bern, J.J.M.~Carrasco, L.J.~Dixon, H.~Johansson and R.~Roiban, \emph{{The
  Complete Four-Loop Four-Point Amplitude in N=4 Super-Yang-Mills Theory}},
  \href{https://doi.org/10.1103/PhysRevD.82.125040}{\emph{Phys. Rev. D}
  {\bfseries 82} (2010) 125040}
  [\href{https://arxiv.org/abs/1008.3327}{{\ttfamily 1008.3327}}].

\bibitem{Bern:2012uc}
Z.~Bern, J.J.M.~Carrasco, H.~Johansson and R.~Roiban, \emph{{The Five-Loop
  Four-Point Amplitude of N=4 super-Yang-Mills Theory}},
  \href{https://doi.org/10.1103/PhysRevLett.109.241602}{\emph{Phys. Rev. Lett.}
  {\bfseries 109} (2012) 241602}
  [\href{https://arxiv.org/abs/1207.6666}{{\ttfamily 1207.6666}}].

\bibitem{Bern:2013uka}
Z.~Bern, S.~Davies, T.~Dennen, A.V.~Smirnov and V.A.~Smirnov,
  \emph{{Ultraviolet Properties of N=4 Supergravity at Four Loops}},
  \href{https://doi.org/10.1103/PhysRevLett.111.231302}{\emph{Phys. Rev. Lett.}
  {\bfseries 111} (2013) 231302}
  [\href{https://arxiv.org/abs/1309.2498}{{\ttfamily 1309.2498}}].

\bibitem{Monteiro:2014cda}
R.~Monteiro, D.~O'Connell and C.D.~White, \emph{{Black holes and the double
  copy}}, \href{https://doi.org/10.1007/JHEP12(2014)056}{\emph{JHEP} {\bfseries
  12} (2014) 056} [\href{https://arxiv.org/abs/1410.0239}{{\ttfamily
  1410.0239}}].

\bibitem{Luna:2015paa}
A.~Luna, R.~Monteiro, D.~O'Connell and C.D.~White, \emph{{The classical double
  copy for Taub\textendash{}NUT spacetime}},
  \href{https://doi.org/10.1016/j.physletb.2015.09.021}{\emph{Phys. Lett. B}
  {\bfseries 750} (2015) 272}
  [\href{https://arxiv.org/abs/1507.01869}{{\ttfamily 1507.01869}}].

\bibitem{Kim:2019jwm}
K.~Kim, K.~Lee, R.~Monteiro, I.~Nicholson and D.~Peinador~Veiga, \emph{{The
  Classical Double Copy of a Point Charge}},
  \href{https://doi.org/10.1007/JHEP02(2020)046}{\emph{JHEP} {\bfseries 02}
  (2020) 046} [\href{https://arxiv.org/abs/1912.02177}{{\ttfamily
  1912.02177}}].

\bibitem{Monteiro:2020plf}
R.~Monteiro, D.~O'Connell, D.~Peinador~Veiga and M.~Sergola, \emph{{Classical
  solutions and their double copy in split signature}},
  \href{https://doi.org/10.1007/JHEP05(2021)268}{\emph{JHEP} {\bfseries 05}
  (2021) 268} [\href{https://arxiv.org/abs/2012.11190}{{\ttfamily
  2012.11190}}].

\bibitem{Crawley:2021auj}
E.~Crawley, A.~Guevara, N.~Miller and A.~Strominger, \emph{{Black holes in
  Klein space}}, \href{https://doi.org/10.1007/JHEP10(2022)135}{\emph{JHEP}
  {\bfseries 10} (2022) 135}
  [\href{https://arxiv.org/abs/2112.03954}{{\ttfamily 2112.03954}}].

\bibitem{Guevara:2021yud}
A.~Guevara, \emph{{Reconstructing Classical Spacetimes from the S-Matrix in
  Twistor Space}},  \href{https://arxiv.org/abs/2112.05111}{{\ttfamily
  2112.05111}}.

\bibitem{Monteiro:2021ztt}
R.~Monteiro, S.~Nagy, D.~O'Connell, D.~Peinador~Veiga and M.~Sergola,
  \emph{{NS-NS spacetimes from amplitudes}},
  \href{https://doi.org/10.1007/JHEP06(2022)021}{\emph{JHEP} {\bfseries 06}
  (2022) 021} [\href{https://arxiv.org/abs/2112.08336}{{\ttfamily
  2112.08336}}].

\bibitem{Luna:2022dxo}
A.~Luna, N.~Moynihan and C.D.~White, \emph{{Why is the Weyl double copy local
  in position space?}},
  \href{https://doi.org/10.1007/JHEP12(2022)046}{\emph{JHEP} {\bfseries 12}
  (2022) 046} [\href{https://arxiv.org/abs/2208.08548}{{\ttfamily
  2208.08548}}].

\bibitem{Carrasco:2015iwa}
J.J.M.~Carrasco, \emph{{Gauge and Gravity Amplitude Relations}},  in
  \emph{{Theoretical Advanced Study Institute in Elementary Particle Physics}:
  {Journeys Through the Precision Frontier: Amplitudes for Colliders}},
  pp.~477--557, WSP, 2015,
  \href{https://doi.org/10.1142/9789814678766_0011}{DOI}
  [\href{https://arxiv.org/abs/1506.00974}{{\ttfamily 1506.00974}}].

\bibitem{Bern:2019prr}
Z.~Bern, J.J.~Carrasco, M.~Chiodaroli, H.~Johansson and R.~Roiban, \emph{{The
  Duality Between Color and Kinematics and its Applications}},
  \href{https://arxiv.org/abs/1909.01358}{{\ttfamily 1909.01358}}.

\bibitem{Bern:2022wqg}
Z.~Bern, J.J.~Carrasco, M.~Chiodaroli, H.~Johansson and R.~Roiban, \emph{{The
  SAGEX review on scattering amplitudes Chapter 2: An invitation to
  color-kinematics duality and the double copy}},
  \href{https://doi.org/10.1088/1751-8121/ac93cf}{\emph{J. Phys. A} {\bfseries
  55} (2022) 443003} [\href{https://arxiv.org/abs/2203.13013}{{\ttfamily
  2203.13013}}].

\bibitem{Kosower:2022yvp}
D.A.~Kosower, R.~Monteiro and D.~O'Connell, \emph{{The SAGEX review on
  scattering amplitudes Chapter 14: Classical gravity from scattering
  amplitudes}}, \href{https://doi.org/10.1088/1751-8121/ac8846}{\emph{J. Phys.
  A} {\bfseries 55} (2022) 443015}
  [\href{https://arxiv.org/abs/2203.13025}{{\ttfamily 2203.13025}}].

\bibitem{Adamo:2022dcm}
T.~Adamo, J.J.M.~Carrasco, M.~Carrillo-Gonz\'alez, M.~Chiodaroli, H.~Elvang,
  H.~Johansson et~al., \emph{{Snowmass White Paper: the Double Copy and its
  Applications}},  in \emph{{Snowmass 2021}}, 4, 2022
  [\href{https://arxiv.org/abs/2204.06547}{{\ttfamily 2204.06547}}].

\bibitem{Luna:2016hge}
A.~Luna, R.~Monteiro, I.~Nicholson, A.~Ochirov, D.~O'Connell, N.~Westerberg
  et~al., \emph{{Perturbative spacetimes from Yang-Mills theory}},
  \href{https://doi.org/10.1007/JHEP04(2017)069}{\emph{JHEP} {\bfseries 04}
  (2017) 069} [\href{https://arxiv.org/abs/1611.07508}{{\ttfamily
  1611.07508}}].

\bibitem{Guevara:2020xjx}
A.~Guevara, B.~Maybee, A.~Ochirov, D.~O'Connell and J.~Vines, \emph{{A
  worldsheet for Kerr}},
  \href{https://doi.org/10.1007/JHEP03(2021)201}{\emph{JHEP} {\bfseries 03}
  (2021) 201} [\href{https://arxiv.org/abs/2012.11570}{{\ttfamily
  2012.11570}}].

\bibitem{Bern:2019crd}
Z.~Bern, C.~Cheung, R.~Roiban, C.-H.~Shen, M.P.~Solon and M.~Zeng, \emph{{Black
  Hole Binary Dynamics from the Double Copy and Effective Theory}},
  \href{https://doi.org/10.1007/JHEP10(2019)206}{\emph{JHEP} {\bfseries 10}
  (2019) 206} [\href{https://arxiv.org/abs/1908.01493}{{\ttfamily
  1908.01493}}].

\bibitem{Bern:2021yeh}
Z.~Bern, J.~Parra-Martinez, R.~Roiban, M.S.~Ruf, C.-H.~Shen, M.P.~Solon et~al.,
  \emph{{Scattering Amplitudes, the Tail Effect, and Conservative Binary
  Dynamics at O(G4)}},
  \href{https://doi.org/10.1103/PhysRevLett.128.161103}{\emph{Phys. Rev. Lett.}
  {\bfseries 128} (2022) 161103}
  [\href{https://arxiv.org/abs/2112.10750}{{\ttfamily 2112.10750}}].

\bibitem{Kosower:2018adc}
D.A.~Kosower, B.~Maybee and D.~O'Connell, \emph{{Amplitudes, Observables, and
  Classical Scattering}},
  \href{https://doi.org/10.1007/JHEP02(2019)137}{\emph{JHEP} {\bfseries 02}
  (2019) 137} [\href{https://arxiv.org/abs/1811.10950}{{\ttfamily
  1811.10950}}].

\bibitem{Maybee:2019jus}
B.~Maybee, D.~O'Connell and J.~Vines, \emph{{Observables and amplitudes for
  spinning particles and black holes}},
  \href{https://doi.org/10.1007/JHEP12(2019)156}{\emph{JHEP} {\bfseries 12}
  (2019) 156} [\href{https://arxiv.org/abs/1906.09260}{{\ttfamily
  1906.09260}}].

\bibitem{delaCruz:2020bbn}
L.~de~la Cruz, B.~Maybee, D.~O'Connell and A.~Ross, \emph{{Classical Yang-Mills
  observables from amplitudes}},
  \href{https://doi.org/10.1007/JHEP12(2020)076}{\emph{JHEP} {\bfseries 12}
  (2020) 076} [\href{https://arxiv.org/abs/2009.03842}{{\ttfamily
  2009.03842}}].

\bibitem{Cristofoli:2021vyo}
A.~Cristofoli, R.~Gonzo, D.A.~Kosower and D.~O'Connell, \emph{{Waveforms from
  amplitudes}}, \href{https://doi.org/10.1103/PhysRevD.106.056007}{\emph{Phys.
  Rev. D} {\bfseries 106} (2022) 056007}
  [\href{https://arxiv.org/abs/2107.10193}{{\ttfamily 2107.10193}}].

\bibitem{Aoude:2021oqj}
R.~Aoude and A.~Ochirov, \emph{{Classical observables from coherent-spin
  amplitudes}}, \href{https://doi.org/10.1007/JHEP10(2021)008}{\emph{JHEP}
  {\bfseries 10} (2021) 008}
  [\href{https://arxiv.org/abs/2108.01649}{{\ttfamily 2108.01649}}].

\bibitem{Goldberger:2004jt}
W.D.~Goldberger and I.Z.~Rothstein, \emph{{An Effective field theory of gravity
  for extended objects}},
  \href{https://doi.org/10.1103/PhysRevD.73.104029}{\emph{Phys. Rev. D}
  {\bfseries 73} (2006) 104029}
  [\href{https://arxiv.org/abs/hep-th/0409156}{{\ttfamily hep-th/0409156}}].

\bibitem{Neill:2013wsa}
D.~Neill and I.Z.~Rothstein, \emph{{Classical Space-Times from the S Matrix}},
  \href{https://doi.org/10.1016/j.nuclphysb.2013.09.007}{\emph{Nucl. Phys. B}
  {\bfseries 877} (2013) 177}
  [\href{https://arxiv.org/abs/1304.7263}{{\ttfamily 1304.7263}}].

\bibitem{Porto:2016pyg}
R.A.~Porto, \emph{{The effective field theorist\textquoteright{}s approach to
  gravitational dynamics}},
  \href{https://doi.org/10.1016/j.physrep.2016.04.003}{\emph{Phys. Rept.}
  {\bfseries 633} (2016) 1} [\href{https://arxiv.org/abs/1601.04914}{{\ttfamily
  1601.04914}}].

\bibitem{Mogull:2020sak}
G.~Mogull, J.~Plefka and J.~Steinhoff, \emph{{Classical black hole scattering
  from a worldline quantum field theory}},
  \href{https://doi.org/10.1007/JHEP02(2021)048}{\emph{JHEP} {\bfseries 02}
  (2021) 048} [\href{https://arxiv.org/abs/2010.02865}{{\ttfamily
  2010.02865}}].

\bibitem{Jakobsen:2021zvh}
G.U.~Jakobsen, G.~Mogull, J.~Plefka and J.~Steinhoff, \emph{{SUSY in the sky
  with gravitons}}, \href{https://doi.org/10.1007/JHEP01(2022)027}{\emph{JHEP}
  {\bfseries 01} (2022) 027}
  [\href{https://arxiv.org/abs/2109.04465}{{\ttfamily 2109.04465}}].

\bibitem{Goldberger:2016iau}
W.D.~Goldberger and A.K.~Ridgway, \emph{{Radiation and the classical double
  copy for color charges}},
  \href{https://doi.org/10.1103/PhysRevD.95.125010}{\emph{Phys. Rev. D}
  {\bfseries 95} (2017) 125010}
  [\href{https://arxiv.org/abs/1611.03493}{{\ttfamily 1611.03493}}].

\bibitem{Goldberger:2017frp}
W.D.~Goldberger, S.G.~Prabhu and J.O.~Thompson, \emph{{Classical gluon and
  graviton radiation from the bi-adjoint scalar double copy}},
  \href{https://doi.org/10.1103/PhysRevD.96.065009}{\emph{Phys. Rev. D}
  {\bfseries 96} (2017) 065009}
  [\href{https://arxiv.org/abs/1705.09263}{{\ttfamily 1705.09263}}].

\bibitem{Luna:2017dtq}
A.~Luna, I.~Nicholson, D.~O'Connell and C.D.~White, \emph{{Inelastic Black Hole
  Scattering from Charged Scalar Amplitudes}},
  \href{https://doi.org/10.1007/JHEP03(2018)044}{\emph{JHEP} {\bfseries 03}
  (2018) 044} [\href{https://arxiv.org/abs/1711.03901}{{\ttfamily
  1711.03901}}].

\bibitem{Goldberger:2017vcg}
W.D.~Goldberger and A.K.~Ridgway, \emph{{Bound states and the classical double
  copy}}, \href{https://doi.org/10.1103/PhysRevD.97.085019}{\emph{Phys. Rev. D}
  {\bfseries 97} (2018) 085019}
  [\href{https://arxiv.org/abs/1711.09493}{{\ttfamily 1711.09493}}].

\bibitem{Goldberger:2017ogt}
W.D.~Goldberger, J.~Li and S.G.~Prabhu, \emph{{Spinning particles, axion
  radiation, and the classical double copy}},
  \href{https://doi.org/10.1103/PhysRevD.97.105018}{\emph{Phys. Rev. D}
  {\bfseries 97} (2018) 105018}
  [\href{https://arxiv.org/abs/1712.09250}{{\ttfamily 1712.09250}}].

\bibitem{Chester:2017vcz}
D.~Chester, \emph{{Radiative double copy for Einstein-Yang-Mills theory}},
  \href{https://doi.org/10.1103/PhysRevD.97.084025}{\emph{Phys. Rev. D}
  {\bfseries 97} (2018) 084025}
  [\href{https://arxiv.org/abs/1712.08684}{{\ttfamily 1712.08684}}].

\bibitem{Li:2018qap}
J.~Li and S.G.~Prabhu, \emph{{Gravitational radiation from the classical
  spinning double copy}},
  \href{https://doi.org/10.1103/PhysRevD.97.105019}{\emph{Phys. Rev. D}
  {\bfseries 97} (2018) 105019}
  [\href{https://arxiv.org/abs/1803.02405}{{\ttfamily 1803.02405}}].

\bibitem{Shen:2018ebu}
C.-H.~Shen, \emph{{Gravitational Radiation from Color-Kinematics Duality}},
  \href{https://doi.org/10.1007/JHEP11(2018)162}{\emph{JHEP} {\bfseries 11}
  (2018) 162} [\href{https://arxiv.org/abs/1806.07388}{{\ttfamily
  1806.07388}}].

\bibitem{Plefka:2018dpa}
J.~Plefka, J.~Steinhoff and W.~Wormsbecher, \emph{{Effective action of dilaton
  gravity as the classical double copy of Yang-Mills theory}},
  \href{https://doi.org/10.1103/PhysRevD.99.024021}{\emph{Phys. Rev. D}
  {\bfseries 99} (2019) 024021}
  [\href{https://arxiv.org/abs/1807.09859}{{\ttfamily 1807.09859}}].

\bibitem{Plefka:2019hmz}
J.~Plefka, C.~Shi, J.~Steinhoff and T.~Wang, \emph{{Breakdown of the classical
  double copy for the effective action of dilaton-gravity at NNLO}},
  \href{https://doi.org/10.1103/PhysRevD.100.086006}{\emph{Phys. Rev. D}
  {\bfseries 100} (2019) 086006}
  [\href{https://arxiv.org/abs/1906.05875}{{\ttfamily 1906.05875}}].

\bibitem{Bastianelli:2021rbt}
F.~Bastianelli, F.~Comberiati and L.~de~la Cruz, \emph{{Worldline description
  of a bi-adjoint scalar and the zeroth copy}},
  \href{https://doi.org/10.1007/JHEP12(2021)023}{\emph{JHEP} {\bfseries 12}
  (2021) 023} [\href{https://arxiv.org/abs/2107.10130}{{\ttfamily
  2107.10130}}].

\bibitem{Shi:2021qsb}
C.~Shi and J.~Plefka, \emph{{Classical double copy of worldline quantum field
  theory}}, \href{https://doi.org/10.1103/PhysRevD.105.026007}{\emph{Phys. Rev.
  D} {\bfseries 105} (2022) 026007}
  [\href{https://arxiv.org/abs/2109.10345}{{\ttfamily 2109.10345}}].

\bibitem{Gonzo:2021drq}
R.~Gonzo and C.~Shi, \emph{{Geodesics from classical double copy}},
  \href{https://doi.org/10.1103/PhysRevD.104.105012}{\emph{Phys. Rev. D}
  {\bfseries 104} (2021) 105012}
  [\href{https://arxiv.org/abs/2109.01072}{{\ttfamily 2109.01072}}].

\bibitem{Ball:2023xnr}
A.~Ball, A.~Bencke, Y.~Chen and A.~Volovich, \emph{{Hidden symmetry in the
  double copy}}, \href{https://doi.org/10.1007/JHEP10(2023)085}{\emph{JHEP}
  {\bfseries 10} (2023) 085}
  [\href{https://arxiv.org/abs/2307.01338}{{\ttfamily 2307.01338}}].

\bibitem{Plebanski:1975wn}
J.F.~Plebanski, \emph{{Some solutions of complex Einstein equations}},
  \href{https://doi.org/10.1063/1.522505}{\emph{J. Math. Phys.} {\bfseries 16}
  (1975) 2395}.

\bibitem{Park:1990fp}
Q.-H.~Park, \emph{{Selfdual Yang-Mills (+ gravity) as a 2-D sigma model}},
  \href{https://doi.org/10.1016/0370-2693(91)90866-O}{\emph{Phys. Lett. B}
  {\bfseries 257} (1991) 105}.

\bibitem{Siegel:1992wd}
W.~Siegel, \emph{{Selfdual N=8 supergravity as closed N=2 (N=4) strings}},
  \href{https://doi.org/10.1103/PhysRevD.47.2504}{\emph{Phys. Rev. D}
  {\bfseries 47} (1993) 2504}
  [\href{https://arxiv.org/abs/hep-th/9207043}{{\ttfamily hep-th/9207043}}].

\bibitem{Bardeen:1995gk}
W.A.~Bardeen, \emph{{Selfdual Yang-Mills theory, integrability and multiparton
  amplitudes}}, \href{https://doi.org/10.1143/PTPS.123.1}{\emph{Prog. Theor.
  Phys. Suppl.} {\bfseries 123} (1996) 1}.

\bibitem{Cangemi:1996rx}
D.~Cangemi, \emph{{Selfdual Yang-Mills theory and one loop like - helicity QCD
  multi - gluon amplitudes}},
  \href{https://doi.org/10.1016/S0550-3213(96)00586-X}{\emph{Nucl. Phys. B}
  {\bfseries 484} (1997) 521}
  [\href{https://arxiv.org/abs/hep-th/9605208}{{\ttfamily hep-th/9605208}}].

\bibitem{Chalmers:1996rq}
G.~Chalmers and W.~Siegel, \emph{{The Selfdual sector of QCD amplitudes}},
  \href{https://doi.org/10.1103/PhysRevD.54.7628}{\emph{Phys. Rev. D}
  {\bfseries 54} (1996) 7628}
  [\href{https://arxiv.org/abs/hep-th/9606061}{{\ttfamily hep-th/9606061}}].

\bibitem{Monteiro:2011pc}
R.~Monteiro and D.~O'Connell, \emph{{The Kinematic Algebra From the Self-Dual
  Sector}}, \href{https://doi.org/10.1007/JHEP07(2011)007}{\emph{JHEP}
  {\bfseries 07} (2011) 007} [\href{https://arxiv.org/abs/1105.2565}{{\ttfamily
  1105.2565}}].

\bibitem{Boels:2013bi}
R.H.~Boels, R.S.~Isermann, R.~Monteiro and D.~O'Connell,
  \emph{{Colour-Kinematics Duality for One-Loop Rational Amplitudes}},
  \href{https://doi.org/10.1007/JHEP04(2013)107}{\emph{JHEP} {\bfseries 04}
  (2013) 107} [\href{https://arxiv.org/abs/1301.4165}{{\ttfamily 1301.4165}}].

\bibitem{Armstrong-Williams:2022apo}
K.~Armstrong-Williams, C.D.~White and S.~Wikeley, \emph{{Non-perturbative
  aspects of the self-dual double copy}},
  \href{https://doi.org/10.1007/JHEP08(2022)160}{\emph{JHEP} {\bfseries 08}
  (2022) 160} [\href{https://arxiv.org/abs/2205.02136}{{\ttfamily
  2205.02136}}].

\bibitem{Monteiro:2022nqt}
R.~Monteiro, R.~Stark-Much\~ao and S.~Wikeley, \emph{{Anomaly and double copy
  in quantum self-dual Yang-Mills and gravity}},
  \href{https://doi.org/10.1007/JHEP09(2023)030}{\emph{JHEP} {\bfseries 09}
  (2023) 030} [\href{https://arxiv.org/abs/2211.12407}{{\ttfamily
  2211.12407}}].

\bibitem{Doran:2023cmj}
G.~Doran, R.~Monteiro and S.~Wikeley, \emph{{On the anomaly interpretation of
  amplitudes in self-dual Yang-Mills and gravity}},
  \href{https://arxiv.org/abs/2312.13267}{{\ttfamily 2312.13267}}.

\bibitem{Balachandran:1976ya}
A.P.~Balachandran, P.~Salomonson, B.-S.~Skagerstam and J.-O.~Winnberg,
  \emph{{Classical Description of Particle Interacting with Nonabelian Gauge
  Field}}, \href{https://doi.org/10.1103/PhysRevD.15.2308}{\emph{Phys. Rev. D}
  {\bfseries 15} (1977) 2308}.

\bibitem{Sternberg:1977rf}
S.~Sternberg, \emph{{Minimal Coupling and the Symplectic Mechanics of a
  Classical Particle in the Presence of a Yang-Mills Field}},
  \href{https://doi.org/10.1073/pnas.74.12.5253}{\emph{Proc. Nat. Acad. Sci.}
  {\bfseries 74} (1977) 5253}.

\bibitem{Balachandran:1977ub}
A.P.~Balachandran, S.~Borchardt and A.~Stern, \emph{{Lagrangian and Hamiltonian
  Descriptions of Yang-Mills Particles}},
  \href{https://doi.org/10.1103/PhysRevD.17.3247}{\emph{Phys. Rev. D}
  {\bfseries 17} (1978) 3247}.

\bibitem{Campiglia:2021srh}
M.~Campiglia and S.~Nagy, \emph{{A double copy for asymptotic symmetries in the
  self-dual sector}},
  \href{https://doi.org/10.1007/JHEP03(2021)262}{\emph{JHEP} {\bfseries 03}
  (2021) 262} [\href{https://arxiv.org/abs/2102.01680}{{\ttfamily
  2102.01680}}].

\bibitem{Strominger:2021mtt}
A.~Strominger, \emph{{$w_{1+\infty}$ Algebra and the Celestial Sphere: Infinite
  Towers of Soft Graviton, Photon, and Gluon Symmetries}},
  \href{https://doi.org/10.1103/PhysRevLett.127.221601}{\emph{Phys. Rev. Lett.}
  {\bfseries 127} (2021) 221601}
  [\href{https://arxiv.org/abs/2105.14346}{{\ttfamily 2105.14346}}].

\bibitem{Adamo:2021lrv}
T.~Adamo, L.~Mason and A.~Sharma, \emph{{Celestial $w_{1+\infty}$ Symmetries
  from Twistor Space}},
  \href{https://doi.org/10.3842/SIGMA.2022.016}{\emph{SIGMA} {\bfseries 18}
  (2022) 016} [\href{https://arxiv.org/abs/2110.06066}{{\ttfamily
  2110.06066}}].

\bibitem{Monteiro:2022lwm}
R.~Monteiro, \emph{{Celestial chiral algebras, colour-kinematics duality and
  integrability}}, \href{https://doi.org/10.1007/JHEP01(2023)092}{\emph{JHEP}
  {\bfseries 01} (2023) 092}
  [\href{https://arxiv.org/abs/2208.11179}{{\ttfamily 2208.11179}}].

\bibitem{Chiodaroli:2013upa}
M.~Chiodaroli, Q.~Jin and R.~Roiban, \emph{{Color/kinematics duality for
  general abelian orbifolds of N=4 super Yang-Mills theory}},
  \href{https://doi.org/10.1007/JHEP01(2014)152}{\emph{JHEP} {\bfseries 01}
  (2014) 152} [\href{https://arxiv.org/abs/1311.3600}{{\ttfamily 1311.3600}}].

\bibitem{Johansson:2014zca}
H.~Johansson and A.~Ochirov, \emph{{Pure Gravities via Color-Kinematics Duality
  for Fundamental Matter}},
  \href{https://doi.org/10.1007/JHEP11(2015)046}{\emph{JHEP} {\bfseries 11}
  (2015) 046} [\href{https://arxiv.org/abs/1407.4772}{{\ttfamily 1407.4772}}].

\bibitem{Chiodaroli:2015rdg}
M.~Chiodaroli, M.~Gunaydin, H.~Johansson and R.~Roiban, \emph{{Spontaneously
  Broken Yang-Mills-Einstein Supergravities as Double Copies}},
  \href{https://doi.org/10.1007/JHEP06(2017)064}{\emph{JHEP} {\bfseries 06}
  (2017) 064} [\href{https://arxiv.org/abs/1511.01740}{{\ttfamily
  1511.01740}}].

\bibitem{Ponomarev:2017nrr}
D.~Ponomarev, \emph{{Chiral Higher Spin Theories and Self-Duality}},
  \href{https://doi.org/10.1007/JHEP12(2017)141}{\emph{JHEP} {\bfseries 12}
  (2017) 141} [\href{https://arxiv.org/abs/1710.00270}{{\ttfamily
  1710.00270}}].

\bibitem{Perry:1994kp}
R.J.~Perry, \emph{{Hamiltonian light front field theory and quantum
  chromodynamics}},  in \emph{{Hadrons 94 Workshop}}, 7, 1994
  [\href{https://arxiv.org/abs/hep-th/9407056}{{\ttfamily hep-th/9407056}}].

\bibitem{Ivanovskiy:2023aay}
V.~Ivanovskiy and D.~Ponomarev, \emph{{Light-cone formalism for a point
  particle in a higher-spin background}},
  \href{https://doi.org/10.1007/JHEP09(2023)014}{\emph{JHEP} {\bfseries 09}
  (2023) 014} [\href{https://arxiv.org/abs/2306.13441}{{\ttfamily
  2306.13441}}].

\bibitem{Metsaev:1991mt}
R.R.~Metsaev, \emph{{Poincare invariant dynamics of massless higher spins:
  Fourth order analysis on mass shell}},
  \href{https://doi.org/10.1142/S0217732391000348}{\emph{Mod. Phys. Lett. A}
  {\bfseries 6} (1991) 359}.

\bibitem{Ponomarev:2016lrm}
D.~Ponomarev and E.D.~Skvortsov, \emph{{Light-Front Higher-Spin Theories in
  Flat Space}}, \href{https://doi.org/10.1088/1751-8121/aa56e7}{\emph{J. Phys.
  A} {\bfseries 50} (2017) 095401}
  [\href{https://arxiv.org/abs/1609.04655}{{\ttfamily 1609.04655}}].

\bibitem{Krasnov:2021nsq}
K.~Krasnov, E.~Skvortsov and T.~Tran, \emph{{Actions for self-dual Higher Spin
  Gravities}}, \href{https://doi.org/10.1007/JHEP08(2021)076}{\emph{JHEP}
  {\bfseries 08} (2021) 076}
  [\href{https://arxiv.org/abs/2105.12782}{{\ttfamily 2105.12782}}].

\bibitem{Monteiro:2022xwq}
R.~Monteiro, \emph{{From Moyal deformations to chiral higher-spin theories and
  to celestial algebras}},
  \href{https://doi.org/10.1007/JHEP03(2023)062}{\emph{JHEP} {\bfseries 03}
  (2023) 062} [\href{https://arxiv.org/abs/2212.11266}{{\ttfamily
  2212.11266}}].

\end{thebibliography}\endgroup
\bibliographystyle{JHEP}

\end{document}